\RequirePackage[2020-02-02]{latexrelease}
\documentclass[12pt,preprint]{revtex4}
\usepackage{graphicx}
\usepackage{amsmath}
\usepackage{amsfonts,amssymb}
\usepackage[matrix,arrow,curve,frame,poly,arc]{xy}
\usepackage[english]{babel}

\def\be{\begin{equation}}
\def\ee{\end{equation}}
\def\ba{\begin{eqnarray}}
\def\ea{\end{eqnarray}}

\begin{document}
\bibliographystyle{plainnat}

\title{Exact analytical vacuum solutions of $ R^n $-gravity model depending on two variables}

\author{Maria Shubina }

\email{yurova-m@rambler.ru}

\affiliation{Skobeltsyn Institute of Nuclear Physics\\Lomonosov Moscow State University
\\ Leninskie gory, GSP-1, Moscow 119991, Russian Federation}


\begin{abstract}

In this paper we consider the metric power-law $f(R)\sim R^n $-gravity model for the four-dimensional metric tensor depending on two coordinates. We obtain exact analytical vacuum solutions for different values of $ n $. These solutions contain both non-stationary configurations of the travelling wave type and stationary ones, in particular, depending on one radial variable.

\end{abstract}


\maketitle

\section{Introduction}

Despite the fact that at present the General Theory of Relativity (GR) is the best fundamental theory describing the gravitational interaction there is reason to believe that this theory may be incomplete. General Relativity is based on the Einstein-Hilbert gravitational action with the Lagrange density $ \sqrt{-g} R $, where $ R $ is the Ricci curvature ($ g = \det g_{\mu \nu }$). One of the simplest modification of Einstein's gravity is to consider the Lagrangian as function $ f(R) $, see \cite{SF}-\cite{NO} and references therein, and various forms of this function have appeared in the literature over the past few decades. Among these functions there are quite viable ones that correctly describe the cosmological dynamic, smooth transition between different cosmological eras, correct weak-field limit and dynamics of cosmological perturbations \cite{F2011}. One of the obvious natural modifications is $ f(R) \sim  R^n $ and the choice of such a function was considered by a number of authors, see references below. The opinion about the power-law models is ambiguous. A number of authors have imposed strong limits on the exponent $ n $, namely $ n - 1 \sim 10^{-30} - 10^{-19}$, making  $ R^n $-gravity an unsuitable candidate to generalize General Relativity. So in Ref. \cite{CB2005} Clifton and Barrow used the dynamical systems approach to find the asymptotic attractor of the general solution at large distances. They found the null and time-like geodesics in $ R^{1 + \delta} $-gravity theories. By comparing their results with cosmological and perihelion-precession observations, they established that in Friedmann–Lema\^itre–Robertson–Walker (FLRW) metric the parameter $\delta\sim 10^{ -19}$. Similar limits on $ \delta $ can be found in the works \cite{B2010}-\cite{F2011}. In Ref. \cite{NDGC2011} the propagation of null geodesics is considered for any $f (R)$-gravity and then calculated explicitly for the case of $ R^n $ in static spherically symmetric space-time in order to determine the bending angle. The found results show that despite stronger constraints on $ n $ given by Solar system tests and observation of perihelion precession ($ \delta \sim 10^{-19} $), these constraints on $ n $ may weaken on a cosmological scale. The $\delta \sim 10^{-4}$ constraint was obtained by comparing the calculations with astronomical observations of primordial $D$ and $4He$ abundances \cite{KKKC2015}. Despite the negative opinion about $ R^n $-gravity as a possible candidate for a realistic alternative to dark energy, Faraoni \cite{F2011} believes that "from the mathematical physics point of view, it is perfectly acceptable to study this theory as a toy model in order to obtain analytical or qualitative insight on exact solutions...". Further in this paper two basic theoretical requirements associated with stability for $ f(R) \sim R^{n} $ are considered. First leads to $ f_{RR}(R) \geq 0 $ \cite{DK2003}-\cite{NO2003} which corresponds to $ n \leq 0  $ or $ n \geq 1 $, and second gives an expression for the mass $ m $ of the nontachyonic scalar field $ f_{R}(R) $ as $ m^{2} = \frac{(2-n)}{3(n-1)}R_{0} $, $ R_{0} $ is the de Sitter space Ricci scalar, that gives $ 1 \leq n \leq 2 $ \cite{FN2005}. 

However, despite such limitations models with different values of $n$ continue to be of interest and are considered in the literature. In the review paper \cite{CCT2003} Capozziello \textit{et al} considered a power-law modification $ f(R) \sim R^{n} $ with $ n = -1 $, $ n = \frac{3}{2} $ and the FLRW metric, where for $ n = \frac{3}{2}$ the Lagrangian is conformally equivalent to the Lagrangian of the Liouville field theory. A detailed analysis of dynamics of cosmological models of $ R^n $-gravity is performed in \cite{CDCT2005}. Carloni \textit{et al} used the dynamical systems approach to study the dynamics of $ R^n $-theories in FLRW universes. They found exact solutions to the cosmological equations and analysed their behaviour and stability in terms of the values of the parameter $n$. In Refs. \cite{CB2006}-\cite{CB2006_2} authors have shown that it is possible to find solutions of the Kasner type for $ R^n $-gravity models and that exact Kasner-like solutions do exist for $1/2 \leq n \leq 5/4$  but with different Kasner-index relations to the ones in GR. The case of $ \delta = \frac{1}{4} $ can be seen in \cite{CST2007}. Shear dynamics and isotropisation are discussed in Ref. \cite{LCD2006}. Leach \textit{et al} give the detailed dynamical systems analysis of Bianchi $ I $ cosmologies in $ R^n $-gravity which exhibit local rotational symmetry. They find exact solutions and study their behaviour and stability in terms of the values of the parameter $ n $. An analysis of the vacuum case has established that the phase space contains one isotropic fixed point and a line of fixed points with shear. The isotropic fixed point is a stable node for $ n < \frac{1}{2}$, $ \frac{1}{2}  < n < 1 $ and $ n > \frac{5}{4}$. In the range $ 1 < n < 5/4 $ this point is a unstable node and therefore may be seen as a past attractor. According to the authors the existence of an isotropic past attractor implies that one do not require special initial conditions for inflation to start since the cosmological singularity is FLRW. The dynamics of orthogonal spatially homogeneous Bianchi cosmologies in $ R^n $-gravity is studied in \cite{GLD2007}. Goheer \textit{et al} study the dynamics of orthogonal spatially homogeneous Bianchi cosmologies in $ R^n $-gravity. The authors perform a detailed analysis of the cosmological behaviour in terms of the parameter $ n $ determining all the equilibrium points, their stability and corresponding cosmological evolution. In Ref. \cite{CCT2006} the authors study the behaviour of the deflection angle in the case of power-law Lagrangians; in Ref. \cite{CCT2007} these authors also consider power-law theories of gravity and compute the modified potential and the rotation curve for spherically symmetric and for thin disc mass distributions. The paper \cite{MS2007} presents an analysis of a devised sample of rotation curves in order to check the consequences of a modified $f(R)$ gravity on galactic scales. The authors generalize the results of \cite{CCT2007} and test a wider sample of spirals, improving the analysis methodology for to perform a check of their model on galactic scales in order to investigate its consistency and universality. Further a modified  $ R^n $-gravity which predicts natural unification of early-time inflation with late-time acceleration is proposed in \cite{NO2008}. The evolution of density perturbations of FLRW models in forth order gravity is rigorously analysed in Ref. \cite{CDT2008}. The general perturbations equations are applied to a simple $ R^n $-gravity model and exact solutions of these equations for scales much bigger than the Hubble radius are obtained. The paper \cite{GLD2009} shows that within the class of $ f(R) $-gravity theories exact FLRW power-law perfect fluid solutions exist only for $ f(R) \sim R^n $ even in the nonperturbative non-GR case, while any other $ f(R) $ models can only allow these exact solutions in the GR limit. The vacuum and non-vacuum Bianchi type $ I $ and $ V $ solutions have been considered in \cite{SS2009}-\cite{SS2010}. Six dimensional black holes coupled to Yang–Mills fields have been studied in \cite{MH2011} where an exact solution was found for the case $ f(R) = R^{\frac{1}{2}} $. In Refs. \cite{LS2011}-\cite{L2011} general cosmological scenarios of anisotropic $f (R)$-gravity is constructed. Focusing on $f (R) =  R^n $ ansatz in the case of Kantowski–Sachs geometry authors perform a systematic phase-space and stability analysis and find that at late times the universe can result in a state of accelerating expansion, and additionally, for a particular $ n $ range ($ 2 < n < 3 $), it exhibits phantom behaviour. The results obtained show that anisotropic geometries in modified  gravity theories represent a radically different cosmological behaviour compared to the simple isotropic scenarios. Further using a combination of the dynamical systems approach and the $ 1+3 $ covariant formalism Carloni \textit{et al} \cite{CVF2013} analysed the phase space of Bianchi $ I $ cosmologies filled by a spin fluid in the framework of $f (R)$-gravity with torsion. The authors also considered $ R^{n} $ model because it is is simple and "at the same time preserves most of the most interesting features of the more general models". The paper \cite{JRMKAM2014} discusses new exact solutions for static wormholes in two $f(R)$-models with a noncommutative-geometry background; one of the models is the power-law one with an integer values of the exponent. The authors showed that the basic characteristics remain essentially the same, but the radius of the throat decreases as the power of $f(R)$ increases. In Ref. \cite{MH2016} Mazharimousavi \textit{et al} investigate the field equations of $f(R)$-gravity in the presence of a string cloud and obtain an infinite class of possible metrics that describe black holes or naked singularities in $ 2 + 1$-dimensional $f(R) = R^{n}$-gravity theory with arbitrary $ n $ ($ n \neq \frac{1}{2} $ and $ n \neq \frac{3 \pm \sqrt{5}}{4} $ ). The local isometry between the spacetime of colliding plane waves and black holes solutions in $R^n$-gravity ($n > 1$) is investigated in Ref. \cite{THM2016}. Morris–Thorne static traversable wormhole solutions in different modified theories of gravity, in particular in power-law $f(R) \sim R^n$-gravity are study in Ref. \cite{SB2021}. The paper \cite{SBS2022} presents Casimir spherically symmetric static wormhole solutions for different kinds of $f(R)$-gravity, including $R^n$-model. Sokoliuk \textit{et al} investigate energy conditions and dynamical stability of the wormhole solutions. By using the Casimir energy density and modified Einstein field equations the authors derive suitable shape functions for each modified gravity under consideration. Bajardi \textit{et al} \cite{BABFC2022} checked "the validity of the swampland criteria" in $ f(R) \sim R^n $-gravity and showed that both criteria are
satisfied for $n < 5/3 $ meaning that "$f(R)$ gravity may be helpful also in the ultraviolet regime to overcome issues occurring in the attempts to develop a quantum theory of gravity". We have not covered all the works, especially articles that have appeared more recently, but we can say that the power-law $f(R)$-gravity continue to attract the attention of researchers \cite{{BJCZJ2021}}-\cite{SPGU2022}.

By inspiring with above works, in this article we consider the $f(R) \sim R^n$-gravity models in four-dimensional space-time in which all functions depend on two coordinates. The system of equations of the theory is integrated in quadratures for various sets of theory parameters; for some of them we have obtained exact analytical vacuum solutions for the metric tensor components and scalar curvature. It should be noted that most of the values of $ n $ for which we obtain an explicit form of our solutions are exactly the same as the values of $ n $ which bound the regions of stability in the phase space found in \cite{CDCT2005}-\cite{CB2006_2}, \cite{LCD2006}.

This article is organized as follows. In Section 2 we introduce the notation and present the equations of the model under consideration, which we will solve below. In Section 3 we specify the form of the function $f(R)$ and integration variables. The exact solutions obtained and their analysis are presented in Section 4 and in the Conclusions respectively.

\section{Models under consideration and field equations}

In this paper we consider the metric $f(R)$-gravity model for the case when the metric tensor depends on two variable. The gravitational action without matter fields is 
\be
S = \frac{1}{16 \pi G} \int d^{4}x \sqrt{-\textit{g}} \, f(R),
\ee
where $ G $ is the gravitational constant, $ \textit{g} $ is the determinant of the metric tensor $ g_{\mu\nu} $, $ R = g^{\mu\nu} R_{\mu\nu} $ is the scalar curvature, or the Ricci scalar, $ R_{\mu\nu} $ is the Ricci tensor.
Variation of eq.(1) with respect to the metric gives the field equations \cite{SF}
\be
f_{R}(R) R_{\mu \nu} - \frac{1}{2}f(R)g_{\mu \nu} - [ \nabla_{\mu} \nabla_{\nu} - g_{\mu \nu} \square ] f_{R}(R) =  0,
\ee
where $ f_{R}(R)\equiv \dfrac{df(R)}{dR} $.

We will consider a metric depending on two coordinates, and take the metric interval in the form:
\be
ds^{2} = - 4 F(\zeta, \eta) \, d\zeta d\eta - g_{ab} dx^{a}dx^{b},
\ee
where $ g_{ab} = g_{ab} (\zeta, \eta) $, $ a, b = 1, 2 $. The variables $\zeta$ and $\eta$ can be either one timelike and one spacelike coordinate, or both spacelike. Let us write down equations (2) in these variables. Denote by $ g $ the $ 2 \times 2 $ matrix $ g_{ab} $ of the metric tensor, and let $ \det g_{ab} = \pm \alpha ^{2} $. Then equations (2) will take the form:
\ba
\Big( \alpha  f_{R} M_{, \, \zeta} \, M^{-1}\Big)_{, \, \eta}  +   \Big( \alpha f_{R} M_{, \, \eta} \, M^{-1}\Big)_{, \,\zeta} & = & 0,\,\,\, M = \alpha^{-1}\, g \\
( \alpha  f_{R})_{, \,\, \zeta\eta} - \alpha F (f_{R} R - f) & = & 0 \\
\alpha  f_{R} F R + \alpha {f_{R}}_{, \,\, \zeta\eta} - 2 f_{R} {\alpha}_{, \,\, \zeta\eta} - \frac{1}{2} (\alpha_{, \,\, \eta} {f_{R}}_{, \,\, \zeta}  +  \alpha_{, \,\,\zeta} {f_{R}}_{, \,\,\eta}) & = & 0 
\ea
and two equations for metric coefficient $ F $ are:
\ba
(\ln F)_{, \,\,\zeta} = \dfrac{\ln ( \alpha  f_{R})_{, \,\, \zeta \zeta}}{\ln ( \alpha  f_{R})_{, \,\, \zeta}} - \dfrac{\textit{Tr} \, (g_{, \,\, \zeta} (g^{-1})_{, \,\, \zeta}) - 4 (\ln (f_{R})_{, \,\, \zeta})^{2}}{4 \ln ( \alpha  f_{R})_{, \,\, \zeta}}\\ \nonumber
(\ln F)_{, \,\,\eta} = \dfrac{\ln ( \alpha  f_{R})_{, \,\, \eta\eta}}{\ln ( \alpha  f_{R})_{, \,\, \eta}} - \dfrac{\textit{Tr} \, (g_{, \,\, \eta} (g^{-1})_{, \,\, \eta}) - 4 (\ln (f_{R})_{, \,\, \eta})^{2}}{4 \ln ( \alpha  f_{R})_{, \,\, \eta}}.
\ea
It is also useful to give the expression for $ (\ln F)_{, \,\zeta \,\eta} $:
\be
(\ln F)_{, \,\zeta \,\eta} = \dfrac{\alpha_{, \,\,\zeta} \alpha_{, \,\,\eta}}{\alpha^{2}} + \frac{1}{4} \textit{Tr} \, (g_{, \,\, \zeta} (g^{-1})_{, \,\, \eta}) - \frac{{f_{R}}_{, \,\, \zeta\eta}}{f_{R}} +  \dfrac{\alpha_{, \,\, \eta} {f_{R}}_{, \,\, \zeta}  +  \alpha_{, \,\,\zeta} {f_{R}}_{, \,\,\eta}}{2 \alpha  f_{R}}
\ee
although it is a consequence of the previous equations. 

We parametrize the components of the two-dimensional metric tensor $g_{ab}(\zeta, \eta)$ as follows:
\ba
g_{ab} = \begin{pmatrix}
\psi & \psi \, \omega\\
\psi \, \omega & \psi \, {\omega}^2 + \sigma_{\alpha}^{2} \,\, \alpha^{2}\, \psi^{-1}
\end{pmatrix}, \,\,\, \sigma_{\alpha}^{2} = \pm 1.
\ea

\section{$ f(R) $ and integration variables}

\subsection{The form of the function $ f(R) $ and the chosen ansatz}
 
We will find exact solutions of eqs. (4)-(7) for the function $ f(R) $ of the form
\be
f(R) = C_{R} \, R^{n},\,\,\, C_{R} = const,
\ee
where the power $ n \neq 1 $. Since in our consideration the values of $ n $ are still arbitrary, we will consider the case of constant-sign $ R $, for example, $ R \geq 0 $. It would be possible to set the modulus, but then the function $ f(R) $ becomes non-smooth at zero, and as will be seen from what follows, the metric also becomes non-smooth. 

From eq. (5) we obtain an expression for $ F R $ and substitute it into eq. (6). Eqs. (5)-(6) are identically satisfied on the ansatz:
\be
\alpha(\zeta, \eta) = \sigma_{\alpha} \, {f_{R}}^{\frac{2n-1}{n-2}}, \,\,\, n \neq 2, \,\,\, 
\ee
and then we obtain an expression for $ F $:
\be
F = \frac{n}{n-1} \,  (n  \, C_{R})^{\frac{1}{n-1}} \,\, ({f_{R}}^{\frac{3(n-1)}{n-2}})_{, \,\, \zeta\eta} \,\, {f_{R}}^{-\frac{3n^{2}-5n+1}{(n-1)(n-2)}}.
\ee

Equations (4) and (7)–(8) are not simplified on this ansatz and we will obtain solutions of the system under consideration in terms of combinations of variables $ \zeta $ and $ \eta $ that reduce the partial differential equations (4)-(7) to ordinary differential equations.

\subsection{Choice of integration variables}

In this subsection, we define the variables $ \varpi( \zeta, \eta ) $ and $ \vartheta( \zeta, \eta ) $ that reduce system (4)-(7) to a system of ordinary differential equations and solve it.

First we consider the real "travelling wave" variable $ \varpi \simeq  \zeta + \lambda \eta $. The name "travelling wave" is conditional, since $\zeta $ and $ \eta $ can be expressed in terms of spatial variables only. Let us now specify the type of variables $ \zeta $ and $ \eta $ for two different cases: the $Case \, \textbf{\textit{1}}$ of a non-stationary metric with positive $ \det g_{ab} $ and the $Case \, \textbf{\textit{2}}$ of a metric depending on two spatial coordinates, for example $ \rho $ and $ z $:
\ba
& Case \, \textbf{\textit{1}}:\,\,\,\,&  \zeta = \frac{1}{2} (z + t), \,\, \,\, \eta = \frac{1}{2} (z - t), \,\,\, \sigma_{\alpha}^{2} = + 1\\
& Case \, \textbf{\textit{2}}:\,\,\,\,&  \zeta = \frac{1}{2} (z + i \rho), \,\, \,\, \eta = \frac{1}{2} (z - i \rho), \,\,\, \sigma_{\alpha}^{2} = - 1.
\ea 
Then the metric interval (3) takes the form:
\ba
& Case \, \textbf{\textit{1}}:\,\,\,\,&  ds^{2} =  F(t, z) \, (dt^{2} -  dz^{2}) - g_{ab} dx^{a}dx^{b}\\
& Case \, \textbf{\textit{2}}:\,\,\,\,&  ds^{2} = - F(z, \rho) \, (dz^{2} +  d\rho^{2}) - g_{ab} dx^{a}dx^{b}.
\ea 
The real variable $ \varpi $ is:
\ba
& Case \, \textbf{\textit{1}}:\,\,\,\,& \varpi ( \zeta, \eta ) = \zeta + \lambda \eta =  \frac{1}{1-v} (z - v t), \,\, v =  \frac{\lambda - 1}{\lambda + 1}\\
& Case \, \textbf{\textit{2}}:\,\,\,\,& \varpi( \zeta, \eta ) = e^{ i \arctan{v}} \,\,\,(\zeta + \lambda \eta) = \frac{1}{\sqrt{1 + v^{2}}} \, (z - v \rho),  \,\, v = i \,\, \frac{\lambda - 1}{\lambda + 1}.
\ea 

The second variable $ \vartheta( \zeta, \eta ) $ in which system (4)-(7) is exactly solved is
\be
\vartheta( \zeta, \eta ) = \ln(- \sigma_{\alpha}^{2} \,\, \zeta \,\eta ) = \begin{cases}
Case \, \textbf{\textit{1}} :  &  \ln\frac{t^{2} -  z^{2}}{4} \\
Case \, \textbf{\textit{2}} : &  \ln\frac{z^{2} +  \rho^{2}}{4} 
\end{cases},
\ee
which for $Case \, \textbf{\textit{2}}$ gives the dependence of all functions on the radius-vector $ r $. 

It can be shown that all equations except eq. (5) have the same form in both these variables. Accordingly the solutions for all functions except for $ F $ will have the same form. Therefore in the next subsection we denote both variables as $ \varphi $ and obtain solutions in terms of this $\varphi$ giving separate expressions for $ F(\varpi) $ and  $ F(\vartheta) $.

\section{Exact solution}


For convenience of notation denote 
\be
{f_{R}}^{\frac{3(n-1)}{n-2}} = w. 
\ee 
Matrix equation (4) is equivalent to two equations:
\be
{\omega},_{\varphi}  =  \dfrac{\omega_{0}\,\, w^{\frac{n+1}{3(n-1)}}}{\psi^{2}}, \,\,\, \omega_{0} = const \nonumber
\ee
\be
(\ln \psi)_{, \,\, \varphi \, \varphi} - \dfrac{2n-1}{3(n-1)} \, \dfrac{w_{, \,\, \varphi \, \varphi}}{w} + (\ln w)_{, \,\, \varphi}\, (\ln \psi)_{, \,\, \varphi} -  \sigma_{\alpha}^{2} \dfrac{{\omega_{0}}^{2}}{\psi^{2} w^{\frac{2(n-2)}{3(n-1)}}} = 0.
\ee
The substitution $ \ln \psi = \Psi $ and $ \ln w = W $ brings eq. (21) to the form
\be
\Psi_{, \,\, \varphi \, \varphi} + W_{, \,\, \varphi}\, \Psi_{, \,\, \varphi} - \dfrac{2n-1}{3(n-1)} \, \big( W_{, \,\, \varphi \, \varphi} + (W_{, \,\, \varphi})^{2} \big) - \sigma_{\alpha}^{2}  \omega_{0}^{2} \,\, e^{-2 \Psi -\frac{2(n-2)}{3(n-1)} W} = 0;
\ee
multiplying by $ e^{2W} $ gives the equation
\be
\big((\Psi - \dfrac{2n-1}{3(n-1)}\,W)_{, \,\, \varphi}\big)^{2} = e^{-2W} \, \big(\sigma_{\Psi_{0}} \Psi_{0}^{2} -  \sigma_{\alpha}^{2}  \omega_{0}^{2} e^{-2\Psi + \frac{2(2n-1)}{3(n-1)}\,W} \big),
\ee
where $ \sigma_{\Psi_{0}} = \pm 1 $,  $ \Psi_{0} = $ constant; $ \Psi_{0} $ can be equal to zero only for $ \sigma_{\alpha}^{2} = -1 $. Further integration allows us to obtain the expression for $ \psi $ in terms of $ w $. Consider first the case $ \Psi_{0} \neq 0 $. Introduce the notation:
\be
\Upsilon = \Psi_{0} \int \dfrac{d\varphi}{w} + \tilde{\varphi_{0}}, \, \,\, \tilde{\varphi_{0}} = const.
\ee
Then
\be
\psi = \frac{\omega_{0}}{\Psi_{0}} \, w^{\frac{2n-1}{3(n-1)}} \, \tilde{\psi} \big( \Upsilon \big), 
\ee
and also we obtain the expression for $ \omega $ in terms of $ w $ for $ \omega_{0} \neq 0 $:
\be
\omega = \frac{\Psi_{0}}{\omega_{0}} \, \tilde{\omega} \big(\Upsilon \big) + \tilde{\omega_{0}}, \,\,  \, \tilde{\omega_{0}} = const,
\ee
where 
\ba
\tilde{\psi}(\cdot) = \begin{cases}
\cosh(\cdot),  & \mbox{if } \sigma_{\alpha}^{2} = +1 \\
\sinh(\cdot) , & \mbox{if } \sigma_{\alpha}^{2} = -1, \sigma_{\Psi_{0}} = +1\\
\sin(\cdot) , & \mbox{if } \sigma_{\alpha}^{2} = -1, \sigma_{\Psi_{0}} = -1
\end{cases}, \,\, 
\tilde{\omega}(\cdot) = \begin{cases}
\tanh(\cdot),  & \mbox{if } \sigma_{\alpha}^{2} = +1 \\
\coth (\cdot) , & \mbox{if } \sigma_{\alpha}^{2} = -1,  \sigma_{\Psi_{0}} = +1\\
\cot (\cdot) , & \mbox{if } \sigma_{\alpha}^{2} = -1, \sigma_{\Psi_{0}} = -1
\end{cases}.
\ea

Eq. (12) for $ F $ takes the forms:
\ba
F(\varpi)& =  &\tilde{\lambda} \, \frac{n}{n-1} \, (n  \, C_{R} )^{\frac{1}{n-1}} \,\, w_{, \,\, \varpi \varpi} \,\, {w}^{-\frac{3n^{2}-5n+1}{3(n-1)^{2}}}, \,\,\, 
\tilde{\lambda} = \begin{cases}
\lambda,  & \mbox{if } \sigma_{\alpha}^{2} = +1 \\
1 , & \mbox{if } \sigma_{\alpha}^{2} = -1
\end{cases} \\ 
F(\vartheta) & =  & - \sigma_{\alpha}^{2} \, \frac{n}{n-1} \, (n  \, C_{R} )^{\frac{1}{n-1}} \,\,e^{- \vartheta} \,\, w_{, \,\, \vartheta \vartheta} \,\, {w}^{-\frac{3n^{2}-5n+1}{3(n-1)^{2}}}.
\ea
Using the ansatz (11) and eqs. (25)-(29) we reduce eqs. (7) to the form:
\be
\Big((\ln w_{, \,\, \varphi})_{, \,\, \varphi} \,\, w^{-\frac{2n^{2}-2n-1}{6(n-1)^{2}}}\Big)_{, \,\, \varphi} +  \frac{\Psi_{0}^{2}}{2}\Big((w_{, \,\, \varphi})^{-1}\Big)_{, \,\, \varphi} w^{-\frac{(2n-1)(4n-5)}{6(n-1)^{2}}} = 0.
\ee

Making the substitution $ w_{, \,\, \varphi} = g $ we obtain the equation:
\be
w \, g \, g_{, \, w} - \frac{8n^{2}-14n+5}{6(n-1)^{2}} \,\, g^{2} + A_{0} \, g + \frac{\Psi_{0}^{2}}{2} = 0,
\ee
where $ A_{0} $ is the integration constant. Let us first consider this equation in the general case, when all coefficients are nonzero. Denote $ \frac{8n^{2}-14n+5}{6(n-1)^{2}} = q $. Then the solutions of this equation can be written in quadratures:
\be
w_{\pm} =  w_{0} \exp\Big ({\frac{1}{q}} \int \dfrac{w_{, \,\, \varphi} \, d w_{, \,\, \varphi}}{(w_{, \,\, \varphi} + \frac{A_{0}}{2q})^{2}  \pm B_{0}^{2}}\Big ), \,\, \, \,  \pm  B_{0}^{2} = -\frac{1}{q} \, \bigg(\frac{\Psi_{0}^{2}}{2} + \frac{A_{0}^{2}}{4q}\bigg), \,\, w_{0} = const.
\ee
The integral on the right-hand side is not difficult to take, but the result is a differential equation, which in the general case is difficult to solve exactly. So it is easy to show that
\ba
w_{+} & = & w_{0} \Big((w_{, \,\, \varphi} + \frac{A_{0}}{2q})^{2} + B_{0}^{2} \Big)^{\frac{1}{2q}} \,\, \exp\Big(-\frac{A_{0}}{2|B_{0}|q^{2}} \,\, \arctan \frac{w_{, \,\, \varphi} + \frac{A_{0}}{2q}}{|B_{0}|}\Big), \, 
\frac{\Psi_{0}^{2}}{2q} + \frac{A_{0}^{2}}{4q^{2}} < 0, \nonumber \\
w_{-} & = & w_{0} \Big |w_{, \,\, \varphi} + \frac{A_{0}}{2q} - |B_{0}| \Big |^{\frac{1}{2q} (1 - \frac{A_{0}}{2q|B_{0}|})} \,\, \Big |w_{, \,\, \varphi} + \frac{A_{0}}{2q} + |B_{0}| \Big |^{\frac{1}{2q} (1 + \frac{A_{0}}{2q|B_{0}|})}, \, 
\frac{\Psi_{0}^{2}}{2q} + \frac{A_{0}^{2}}{4q^{2}} > 0,
\ea
however, it seems to us that it is possible to obtain the exact form of the solution $w = w(\varphi)$ only for certain values of the combinations of constants in the equations, and these values are very few. Finding exact solutions for these cases we leave for further study and below we consider exact solutions for simpler cases. Let $ A_{0} = 0 $, then eq. (31) gives: 
\be
\pm \int \dfrac{dw}{\sqrt{C_{0} w^{2q} + \frac{\Psi_{0}^{2}}{2q}}} = \varphi + \varphi_{0}, \,\,\,\,C_{0} = const, \,\, \varphi_{0} = const.
\ee
The integral on the left-hand side is taken in hypergeometric functions, but we are only interested in the solutions that allow us to explicitly express $ w $ and hence all the variables of the theory from $ \varphi $. In these cases the metric interval can be presented in the form:
\ba
ds^{2} & = & - 4 F(\varphi) \, d\zeta d\eta -\, 
\frac{\omega_{0}}{\Psi_{0}} \, w^{\frac{2n-1}{3(n-1)}} \, \tilde{\psi} \big( \Upsilon \big) \, (dx^{1})^{2} \, - \\ \nonumber
& - & \, 2 \,  w^{\frac{2n-1}{3(n-1)}} \,\, \tilde{\psi} \big( \Upsilon \big) \,\, \Big ( \tilde{\omega} \big(\Upsilon \big) + \frac{\tilde{\omega_{0}}\omega_{0} }{\Psi_{0}} \Big) \,\, dx^{1} dx^{2} \\ \nonumber
& - & \frac{\Psi_{0}}{ \omega_{0}} w^{\frac{2n-1}{3(n-1)}} \,\, \Big[ \tilde{\psi} \big( \Upsilon \big) \Big ( \tilde{\omega} \big(\Upsilon \big) + \frac{\tilde{\omega_{0}}\omega_{0} }{\Psi_{0}} \Big)^{2} + \sigma_{\alpha} \tilde{\psi}^{-1} \big( \Upsilon \big) \Big] \, (dx^{2})^{2},
\ea
where we will get exact expressions for $ F $ and $ \Upsilon (\varphi) $ for some values of power $ n $.
Given eq.(27), there are $ 6 $ solutions for each value of $ n $: $ 3 $ for variable $ \varpi $ and $ 3 $ for variable $  \vartheta $. However, not all of these solutions can be considered as components of the metric; not all solutions satisfy the correct physical signature of the metric tensor or some solutions are not of constant sign. So, the non-physical signature is more common in $Case \, \textbf{\textit{2}}$ that is for solutions in variables $ \rho $ and $ z $.

It should also be noted that some solutions change the initial signature of the space-time metric. Since we are interested in solutions in the form of "elementary" functions, it is not difficult to analyze the behaviour of these functions for various values of the argument. But we still schematically present the graphs of some suitable solutions below.

\subsection{$ \Psi_{0} \neq 0 $,  $ A_{0} = 0 $}

\subsubsection{$ n_{\mp} = \dfrac{1 \mp \sqrt{3}}{2}$ }

For the case $ n_{\mp} = \dfrac{1 \mp \sqrt{3}}{2} $ (or $ q = 1 $) the form of the resulting solution depends on the sign of $ C_{0} $ in eq. (34). The exponent of $ w $ in the metric interval is $ \dfrac{2n_{\mp}-1}{3(n_{\mp}-1)}  =  1 \mp \frac{1}{\sqrt{3}}  $.

For $ C_{0} > 0 $ we obtain:
\ba
w & = & \frac{|\Psi_{0}|}{\sqrt{2 C_{0}}} \, \sinh(\sqrt{C_{0}}\varphi + \varphi_{0} ) \\ \nonumber
F & = & F_{0} \, w^{1 \mp \frac{1}{\sqrt{3}}}, \,\,\,\, 
F_{0} = \begin{cases}
\tilde{\lambda} \tilde{F_{0}},  & \mbox{if } \varphi = \varpi \\
- \sigma_{\alpha}^{2} e^{-\vartheta} \, \tilde{F_{0}} , & \mbox{if }  \varphi = \vartheta
\end{cases} \\
\rm{where} \nonumber \\ 
\tilde{F_{0}} & = &  C_{0} (2 \mp \sqrt{3}) \Big (\frac{ C_{R} (2 \mp \sqrt{3})}{1 \mp \sqrt{3}}\Big )^{1 \mp \sqrt{3}} \nonumber \\
\Upsilon & = & \sqrt{2} \ln \Big |\tanh \frac{\sqrt{C_{0}}\varphi + \varphi_{0}}{2} \Big | + \tilde{\varphi_{0}} \nonumber.
\ea
The scalar curvature $ R $ is:
\be
R =  \Big (\frac{1 \mp \sqrt{3}}{C_{R}(2 \mp \sqrt{3})}\Big )^{1 \mp \sqrt{3}} \,\,  w^{-1 \pm \frac{1}{\sqrt{3}}}.
\ee
Schematically the behaviour of metric and $ R $ is shown in Fig.1a - Fig.1d; we take the following constants: $ |C_{R}| = 1 $,  $ |\omega_{0}| = 9 $, $ |\Psi_{0}| = 1 $, $ C_{0} = 4 $ and $ \varphi_{0} = 0 $, $ \lambda  = 3.5 $, $ \tilde{\varphi_{0}} = 1/\sqrt{55} $ in Fig.1a - Fig.1b and $ \tilde{\varphi_{0}} = 0 $ in Fig.1c. Fig.1d shows a spherically symmetric metric with $ C_{0} = 0.04 $,  $ \varphi_{0} = 10 $ and $ \tilde{\varphi_{0}} = -0.14 $. This metric is defined for all $ r > r_{0} $ where $ r_{0} $ can be made arbitrarily small by choosing the constants; in Fig.1d $ r > 5.6 \cdot 10^{-11}$.

\begin{figure}[h!]
\begin{minipage}{0.24\linewidth}
\center{\includegraphics[width=0.8\linewidth]{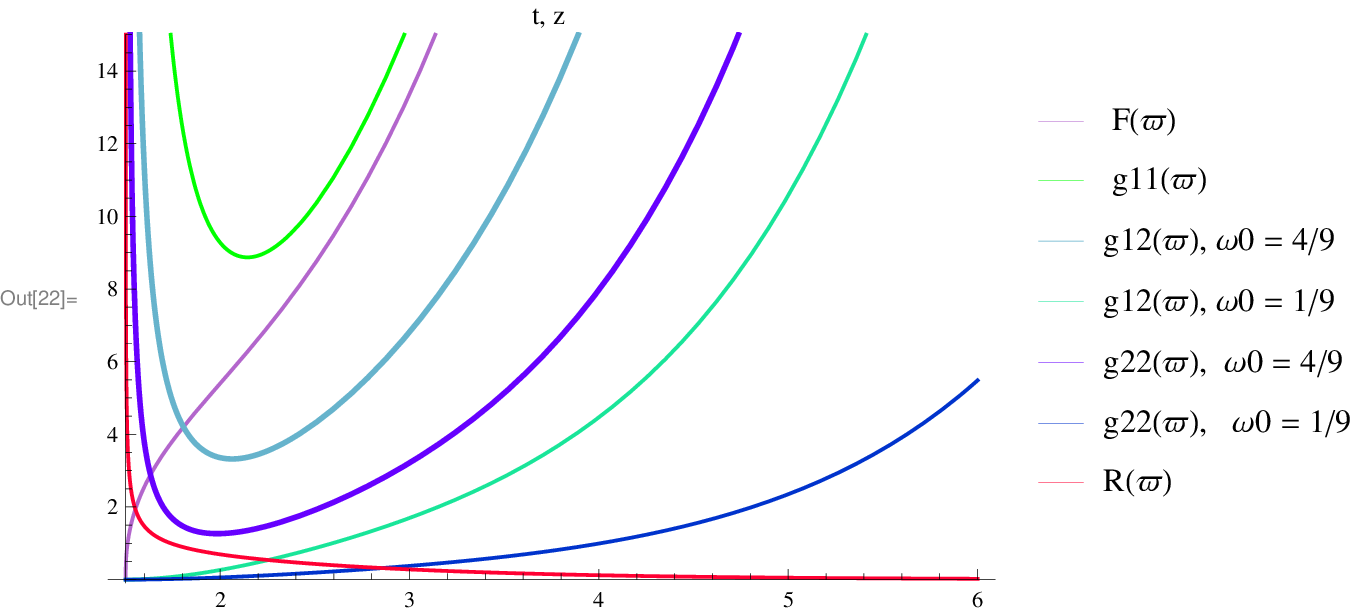} \\ Fig.1a: $  n_{-} = \frac{1 - \sqrt{3}}{2} $;  }
\end{minipage}
\hfill
\begin{minipage}{0.24\linewidth}
\center{\includegraphics[width=0.8\linewidth]{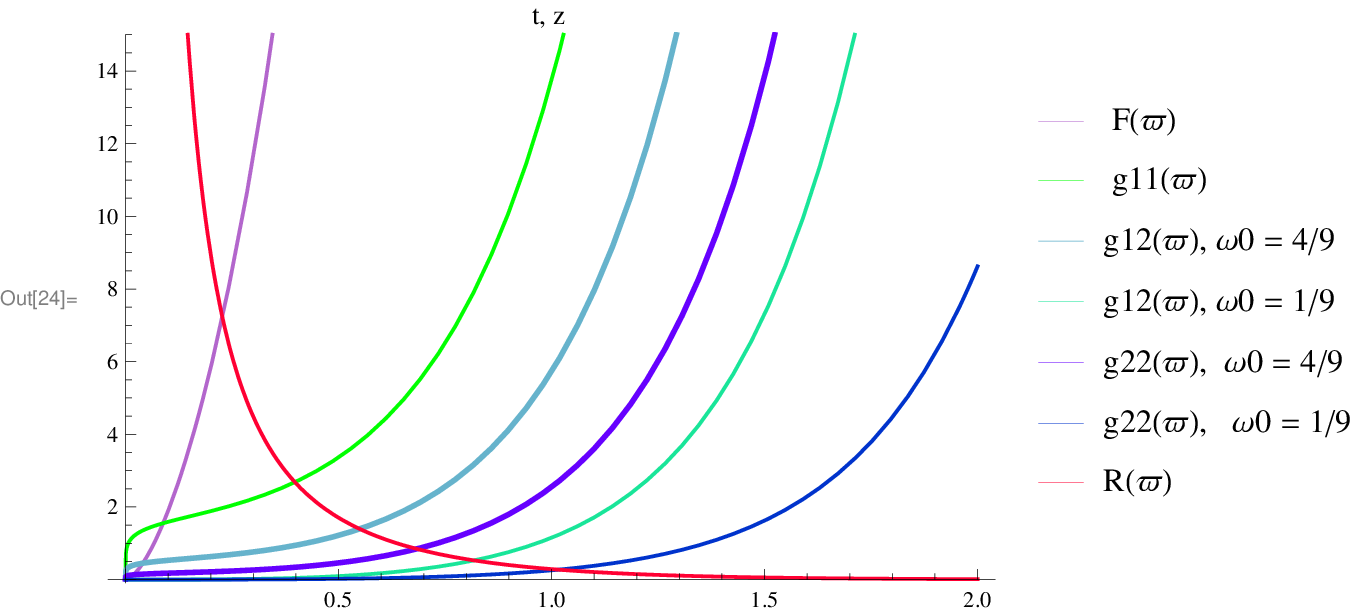} \\ Fig.1b: $ n_{+} = \frac{1 + \sqrt{3}}{2} $;}
\end{minipage}
\hfill
\begin{minipage}{0.24\linewidth}
\center{\includegraphics[width=0.8\linewidth]{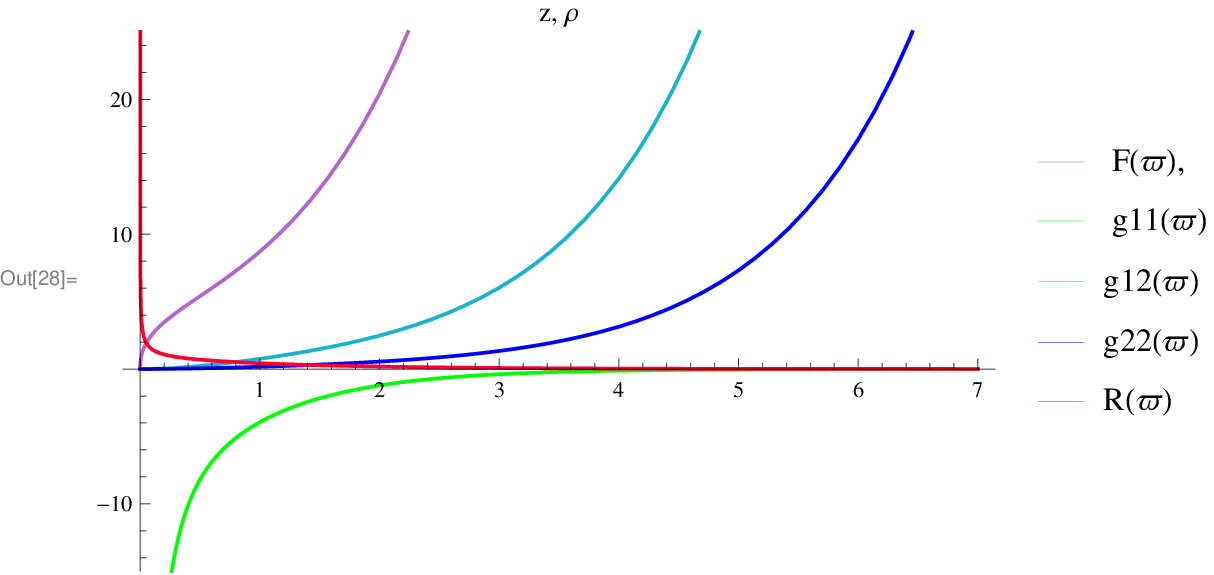} \\ Fig.1c: $ n_{-} = \frac{1 - \sqrt{3}}{2} $; $\sigma_{\Psi_{0}} = +1 $;  $\tilde{\omega_{0}} = \frac{1}{9}$;}
\end{minipage}
\hfill
\begin{minipage}{0.24\linewidth}
\center{\includegraphics[width=0.8\linewidth]{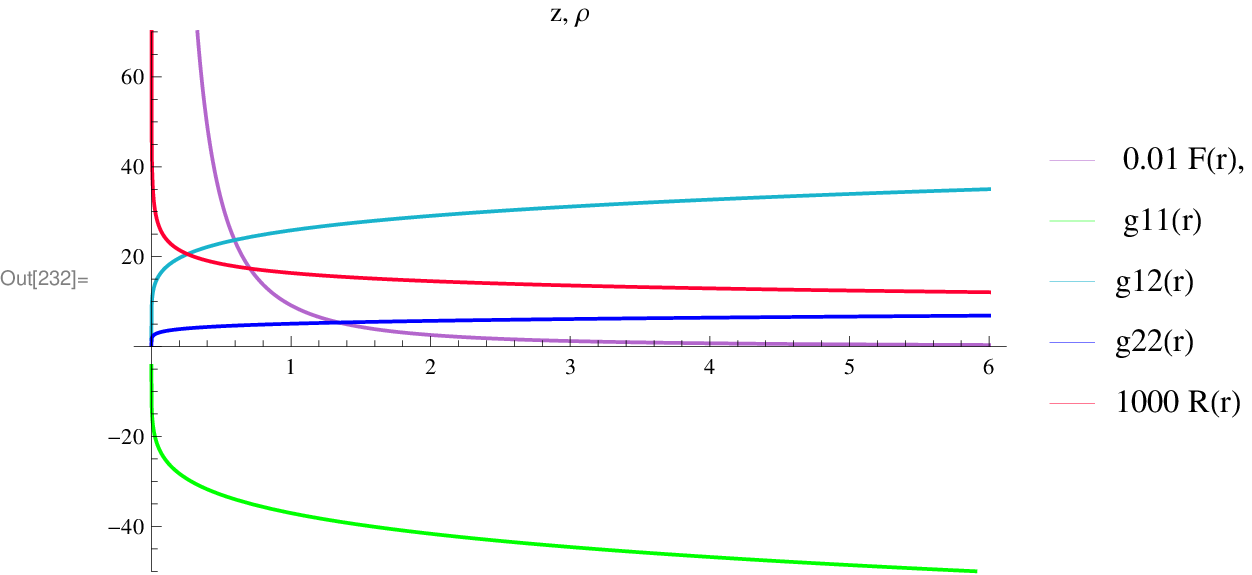} \\ Fig.1d $ n_{-} = \frac{1 - \sqrt{3}}{2} $; $ \sigma_{\Psi_{0}} = +1 $;  $\tilde{\omega_{0}} = \frac{1}{9}$;}
\end{minipage}
\hfill
\end{figure}
Further let $ C_{0} < 0$. Then in eqs. (36)-(37) the hyperbolic functions will be replaced by a trigonometric ones and $C_0 \longrightarrow |C_0| $, $ F_0 \longrightarrow - F_0 $ . 
The resulting solutions do not allow obvious interpretation as components of the metric tensor.


\subsubsection{$n_{\mp} = \dfrac{4 \mp \sqrt{6}}{5} $}

For the case $ n_{\mp} = \dfrac{4 \mp \sqrt{6}}{5} $ (or $ q = \frac{1}{2} $) the exponent of $ w $ in the interval is $ \dfrac{2n_{\mp}-1}{3(n_{\mp}-1)}  =  1 \mp \sqrt{\frac{2}{3}} $. The function $ w$ has the form
\ba
w & = & \frac{1}{C_{0}} \, \Big ((\frac{C_{0}}{2} \varphi + \varphi_{0})^{2} - \Psi_{0}^{2} \Big ) \\ \nonumber
F & = & F_{0} \, w^{1 \mp \sqrt{\frac{2}{3}}},  \\
\rm{where} \nonumber \\ 
\tilde{F_{0}} & = & \frac{C_{0} }{2} (2 \mp \sqrt{6}) \Big (\frac{ C_{R}  (2 \mp \sqrt{6})}{1 \mp \sqrt{6}}\Big )^{1 \mp \sqrt{6}} \nonumber \\
\Upsilon & = &  \ln \Big | \frac{\frac{C_{0}}{2} \varphi + \varphi_{0} - \Psi_{0}}{\frac{C_{0}}{2} \varphi + \varphi_{0} + \Psi_{0}} \Big | + \tilde{\varphi_{0}} \nonumber.
\ea
and $ F_{0} $ is the same as in eq. (36). The scalar curvature $ R $ is:
\be
R =  \Big (\frac{1 \mp \sqrt{6}}{C_{R}(2 \mp \sqrt{6})}\Big )^{1 \mp \sqrt{6}} \,\,  w^{-2 \pm \sqrt{\frac{2}{3}}}.
\ee
Graphs of the a spherically symmetric metric components and $ R $ are shown in Fig.2 for $ |\omega_{0}| = 0.1 $, $ \Psi_{0} = 1 $, $ \tilde{\varphi_{0}} = 2 $, $ |C_{0}| = 2 $, $\tilde{\omega_{0}} = - 10 $ and $ \varphi_{0} = 4 $. From the analysis of eqs. (38)-(39) it can be seen that $ r > r_{0}  $. In Fig.2 for clarity we chose $ r_{0} $ rather large, but $ r_{0} $ can be made arbitrarily small by choosing suitable integration constants.
\begin{figure}[h!]
\begin{minipage}{0.24\linewidth}
\center{\includegraphics[width=0.8\linewidth]{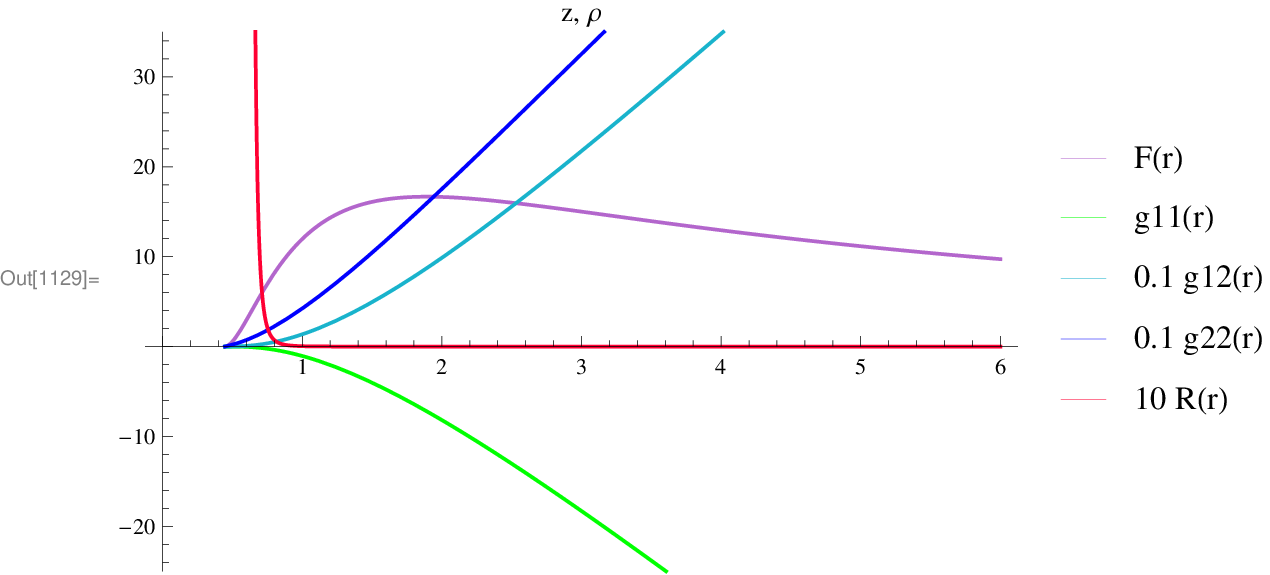} \\ Fig.2: $  n_{+} = \frac{4 + \sqrt{6}}{5} $;  }
\end{minipage}
\hfill
\begin{minipage}{0.24\linewidth}
\center{\includegraphics[width=0.8\linewidth]{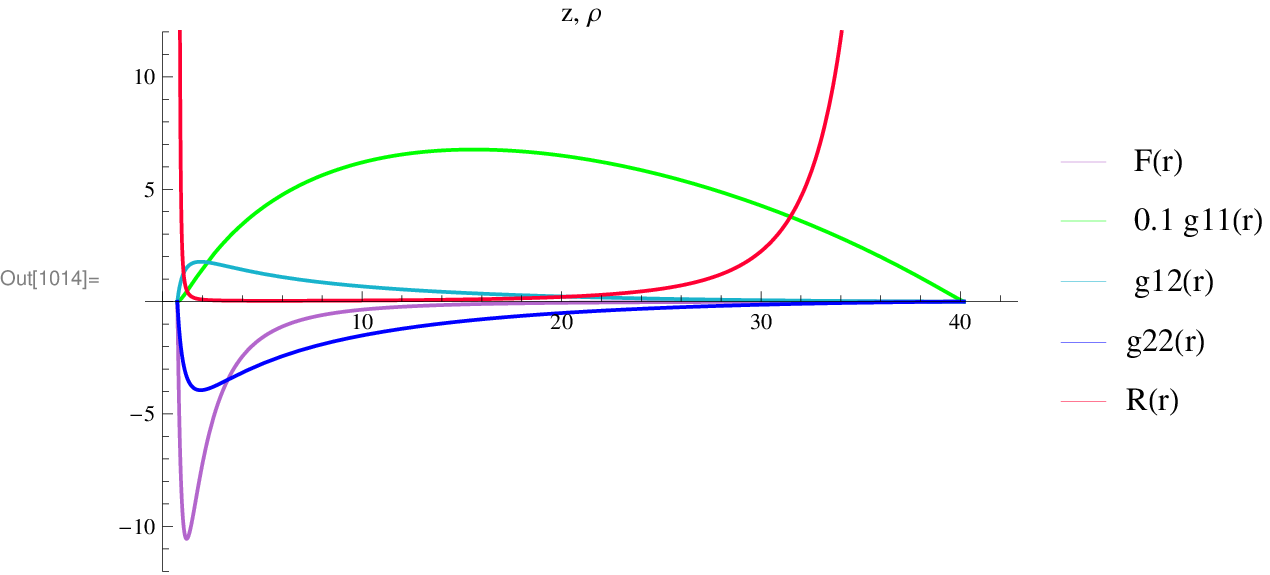} \\ Fig.3: $ n_{+} = \frac{13 + \sqrt{15}}{14} $;}
\end{minipage}
\hfill
\begin{minipage}{0.24\linewidth}
\center{\includegraphics[width=0.8\linewidth]{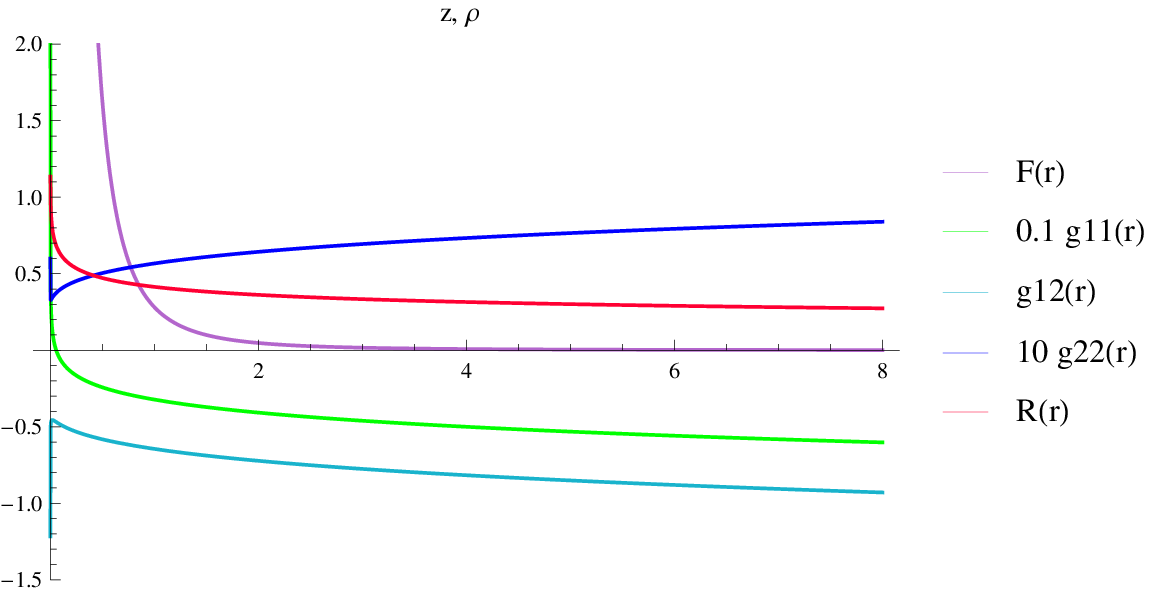} \\ Fig.4: $ n_{-} = \frac{1 - \sqrt{3}}{2} $; }
\end{minipage}
\hfill
\begin{minipage}{0.24\linewidth}
\center{\includegraphics[width=0.8\linewidth]{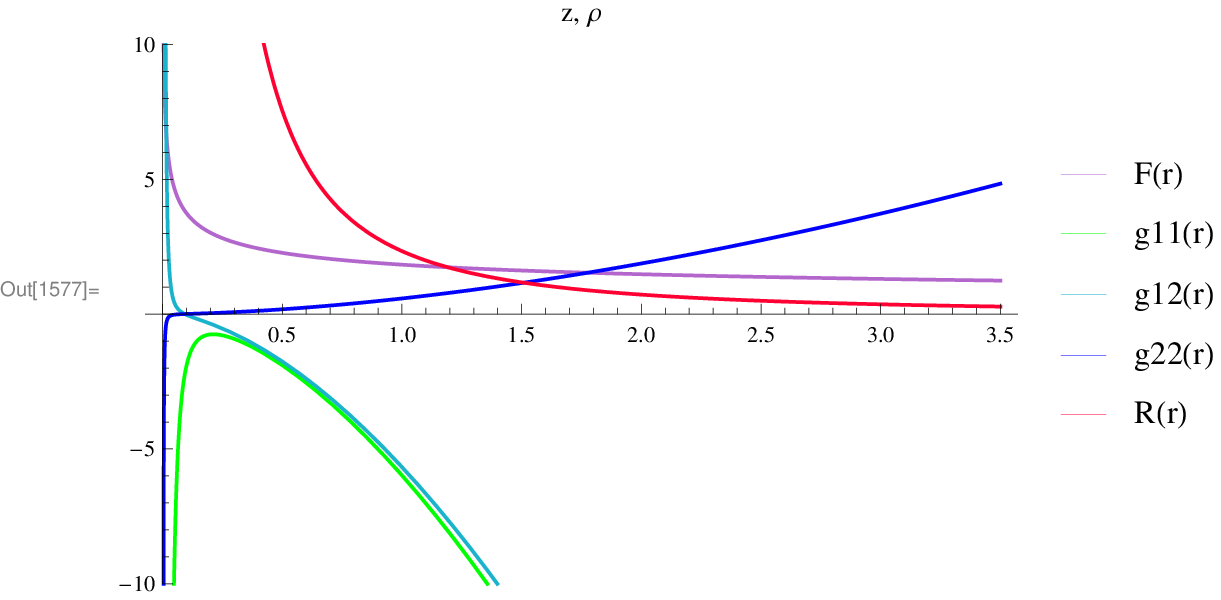} \\ Fig.5 $ n_{-} = \frac{1 - \sqrt{3}}{2} $; }
\end{minipage}
\hfill
\end{figure}

\subsubsection{$n_{\mp} = \dfrac{13 \mp \sqrt{15}}{14} $}

For the case $ n_{\mp} = \dfrac{13 \mp \sqrt{15}}{14} $ (or $ q = -1 $) $ \dfrac{2n_{\mp}-1}{3(n_{\mp}-1)}  =  1 \mp \sqrt{\frac{5}{3}} $. 
The function $ w$ is
\ba
w & = & \frac{\sqrt{2}}{|\Psi_{0}|} \, \sqrt{ C_{0}^{2} -(\frac{\Psi_{0}^{2}}{2} \varphi + \varphi_{0})^{2}} \\ \nonumber 
F & = & F_{0} \, w^{1 \mp \sqrt{\frac{5}{3}}}
\ea
and $ F_{0} $ is the same as in eq. (36);
\ba
\tilde{F_{0}} & = & -  C_{0}^{2} (2 \mp \sqrt{15}) \Big (\frac{ C_{R} (2 \mp \sqrt{15})}{1 \mp \sqrt{15}}\Big )^{1 \mp \sqrt{15}}\\ \nonumber
\Upsilon & = & \sqrt{2} \, \arcsin  \Big ( \frac{1}{|C_{0}|} \big ( \frac{\Psi_{0}^{2}}{2} \varphi + \varphi_{0} \big) \Big ) + \tilde{\varphi_{0}}.
\ea
The scalar curvature $ R $ has the form: 
\be
R =  \Big (\frac{ (1 \mp \sqrt{15})}{C_{R}(2 \mp \sqrt{15})}\Big )^{1 \mp \sqrt{15}} \,\,  w^{\sqrt{\frac{5}{3}} \, (\pm 1 - \sqrt{15})}.
\ee
Eq. (40) shows that $r_{1} < r <  r_{2}  $ for the spherically symmetric case. As in the previous case we have taken $ \Delta r $ small, but $ r_{1}  $ can be made very small and $ r_{2} $ very large by a suitable choice of constants. The components of metric tensor and $ R $ are shown in Fig.3 for $ |\omega_{0}| = 0.9 $, $ \Psi_{0} = 1 $, $ \tilde{\varphi_{0}} = \pi/\sqrt{2} $, $ |C_{0}| = 2 $, $\tilde{\omega_{0}} = - \frac{10}{9}$ and $ \varphi_{0} = -1 $. 

Now consider the case $ A_{0} \neq 0 $.

\subsection{$ \Psi_{0} \neq 0 $,  $A_{0} = \pm \dfrac{\Psi_{0}}{\sqrt{2}} $}

\subsubsection{$ n = \frac{1}{2}$ and $ n = \frac{5}{4}$}

For the case $ n = \frac{1}{2}$ and $ n = \frac{5}{4}$ (or $q = 0$ ) the metric functions are:
\ba
w & = & \frac{|\Psi_{0}|}{\sqrt{2}} \, (\varphi + \varphi_{0}) \,\, \ln |C_{0}(\varphi + \varphi_{0})|   \\ \nonumber
\Upsilon & = & \sqrt{2} \, \ln  \Big | C_{1} \ln \big | \varphi + \varphi_{0} \big| \Big | \,\,\,
\\ \nonumber
F_{\frac{1}{2}} & = & (F_{0})_{\frac{1}{2}} \, \ln |C_{0}(\varphi + \varphi_{0})|, \,\,\, \tilde{(F_{0})_{\frac{1}{2}}} = - \dfrac{2 \Psi_{0}^{2}}{C_{R}^{2}}
\\ \nonumber
F_{\frac{5}{4}} & = & (F_{0})_{\frac{5}{4}} \, (\varphi + \varphi_{0})^{2} \,\, (\ln |C_{0}(\varphi + \varphi_{0})|)^{3}, \,\,\, \tilde{(F_{0})_{\frac{5}{4}}} =  \frac{C_{R}^{4} \Psi_{0}^{4} 5^{5}}{4^{5}} \\ \nonumber
\ea
$ C_{1}= const $.
The scalar curvature $ R $ has the form: 
\ba
R_{\frac{1}{2}} & = & \frac{C_{R}^{2}}{2 \big ( \Psi_{0} (\varphi + \varphi_{0})     \ln |C_{0}(\varphi + \varphi_{0})| \big)^{2}} \\ \nonumber
R_{\frac{5}{4}} & = & \frac{ 4^{5} }{\big ( 5 C_{R}\, \Psi_{0} (\varphi + \varphi_{0})     \ln |C_{0}(\varphi + \varphi_{0})| \big)^{4}}. 
\ea
The resulting solutions do not seem suitable to us. For $Case \, \textbf{\textit{2}}$ as in most of the previous solutions a non-physical signature of the metric tensor is obtained. But even solutions with the correct signature are difficult to interpret.

\subsection{ $\Psi_{0} = 0 $}

Consider now the case $ \Psi_{0} = 0 $, as follows from eq. (23) this is possible only with $ \sigma_{\alpha}^{2} = -1 $. As above, we introduce the notation:
\be
\Upsilon = \omega_{0} \int \dfrac{d\varphi}{w} + \tilde{\varphi_{0}}, \, \,\, \tilde{\varphi_{0}} = const.
\ee
Then
\ba
\psi & = & w^{\frac{2n-1}{3(n-1)}} \,  \Upsilon  \\ \nonumber
\omega & = & - \dfrac{1}{\Upsilon} + \tilde{\omega_{0}}, \,\,  \, \tilde{\omega_{0}} = const.
\ea
Equation (30) is immediately solved and gives:
\be
\int \dfrac{dw}{w^{q} + w_{0}} = C_{0} \varphi + \varphi_{0}, \,\,\, q \neq 0,
\ee
where $ C_{0} $, $ w_{0} $ and $ \varphi_{0} $ are constants. (For $ q = 0 $ the integral is taken in special functions which makes it difficult to express $ w $ in terms of $ \varphi $). For an arbitrary nonzero $ q $ and $ w_{0} $ the integral is expressed in terms of the hypergeometric function, but again we are only interested in solutions that give the explicit form $ w = w(\varphi) $. The metric interval can be presented in the form:
\ba
ds^{2} & = & - 4 F(\varphi) \, d\zeta d\eta -\, 
w^{\frac{2n-1}{3(n-1)}} \, \Upsilon \, (dx^{1})^{2} \, - \\ \nonumber
& - & \, 2 \,  w^{\frac{2n-1}{3(n-1)}} \,\, (\tilde{\omega_{0}} \, \Upsilon - 1)  \,\, dx^{1} dx^{2} 
 -  w^{\frac{2n-1}{3(n-1)}} \,\, \Upsilon^{-1} \,\, \Big[ (\tilde{\omega_{0}} \, \Upsilon - 1)^{2} - 1 \Big] \, (dx^{2})^{2}, \\ \nonumber
F & = &  \tilde{F}\cdot \begin{cases}
1,  & \mbox{if } \varphi = \varpi \\
 e^{-\vartheta}, & \mbox{if }  \varphi = \vartheta.
\end{cases}
\ea

However, among these solutions, very few can be interpreted as metric components. We will present the two most suitable solutions.

\subsubsection{$ n_{\mp} = \dfrac{1 \mp \sqrt{3}}{2}$, $ w_{0} \neq 0 $}

For the case $ n_{\mp} = \dfrac{1 \mp \sqrt{3}}{2} $ (or $ q = 1 $) the form of the solution will obviously differ from eq. (36) and has the form:
\ba
w & = & e^{C_{0} \varphi + \varphi_{0}} \, - w_{0} \nonumber \\ 
\Upsilon & = & - \frac{1}{w_{0} C_{0}} \big( C_{0} \varphi + \varphi_{0} - \ln | e^{C_{0} \varphi + \varphi_{0}} \, - w_{0} |  \big) \nonumber \\ 
\tilde{F} & = & (2 \mp \sqrt{3}) \, \big ( \frac{C_{R} (1 \mp \sqrt{3})}{2}\big )^{\frac{2}{-1 \mp \sqrt{3}}}  \, C_{0}^{2} \, e^{C_{0} \varphi + \varphi_{0}}  \,\,  \big( e^{C_{0} \varphi + \varphi_{0}} \, - w_{0} \big )^{\mp \frac{4}{\sqrt{3}}}\nonumber \\ 
R & = & \big( \frac{C_{R} (1 \mp \sqrt{3})}{2}\big )^\frac{2}{1 \pm \sqrt{3}} \, \Big( e^{C_{0} \varphi + \varphi_{0}} \, - w_{0} \Big )^{-1 \pm \frac{1}{\sqrt{3}}}.
\ea
Fig. 4 shows the spherically symmetric metric for $ |\omega_{0}| = 0.2 $, $ \tilde{\varphi_{0}} = 0 $, $ C_{0} = 0.25 $, $|\tilde{\omega_{0}}| = 0.04$ and $ \varphi_{0} = \ln 0.7 $. As before $ r > r_{0} $ where $ r_{0} $ can be made arbitrarily small.

\subsubsection{$  n_{\mp} = \dfrac{1 \mp \sqrt{3}}{2}$, $ w_{0} = 0 $.}

For $ w_{0} = 0 $ the integral in eq. (47) is taken exactly for any possible $ q $. For $ q = 1 $ the solution has the form:
\ba
w & = & e^{C_{0} \varphi + \varphi_{0}} \nonumber \\ 
\Upsilon & = & - \frac{\omega_{0}}{C_{0}} e^{ - C_{0} \varphi - \varphi_{0} } + \tilde{\varphi_{0}} \nonumber \\ 
\tilde{F} & = & (2 \mp \sqrt{3}) \, \big ( \frac{C_{R} (1 \mp \sqrt{3})}{2}\big )^{\frac{2}{-1 \mp \sqrt{3}}}  \, C_{0}^{2} \,  \big( e^{C_{0} \varphi + \varphi_{0}} \big )^{1 \mp \frac{1}{\sqrt{3}}}\nonumber \\ 
R & = & \big( \frac{C_{R} (1 \mp \sqrt{3})}{2}\big )^\frac{2}{1 \pm \sqrt{3}} \, ( e^{C_{0} \varphi + \varphi_{0}})^{-1 \pm \frac{1}{\sqrt{3}}}.
\ea
On Fig. 5 is shown the spherically symmetric metric for $ |\omega_{0}| = 0.2 $, $ |\tilde{\varphi_{0}}| = 1 $, $ C_{0} = 2 $, $|\tilde{\omega_{0}}| = 0.05$ and $ \varphi_{0} = 7 $. As before the correct metric signature is for $ r > r_{0} $ where $ r_{0} $ can be made arbitrarily small.

\subsubsection{$  n_{\mp} \neq \dfrac{1 \mp \sqrt{3}}{2}$, $ \frac{1}{2} $, $ \frac{5}{4} $;  $ w_{0} = 0 $.}

For $ q \neq 0 $ and $ 1 $ the solutions obtained have the form:
\ba
w & = & (C_{0} \varphi + \varphi_{0})^{- \frac{6(n-1)^{2}}{1+2n-2n^{2}}} \nonumber \\ 
\Upsilon & = & - \frac{\omega_{0} (1+2n-2n^{2})}{(2n-1)(4n-5)} \,\, 
(C_{0} \varphi + \varphi_{0})^{- \frac{(2n-1)(4n-5)}{1+2n-2n^{2}}} \, + \tilde{\varphi_{0}}\nonumber \\ 
\tilde{F} & = & (C_{R}n)^{\frac{1}{n-1}} \,\frac{6n(n-1)(2n-1)(4n-5) C_{0}^{2}}{(1+2n-2n^{2})^{2}} \,\, (C_{0} \varphi + \varphi_{0})^{\frac{2(n-1)(2n-1)}{1+2n-2n^{2}}} \nonumber \\ 
R & = & (C_{R}n)^{\frac{1}{1-n}} \, (C_{0} \varphi + \varphi_{0})^{\frac{2(n-2)}{1+2n-2n^{2}}}.
\ea
The properties of the obtained solutions depend on $ n $, but some general features can be established. Consider the spherically symmetric solutions that preserve the initial signature of the metric $ (+ - - - ) $. In this case the components of the metric tensor must be $ F > 0 $, $ \psi < 0 $ and $ \psi {\omega}^2 - \alpha^{2} \psi^{-1}  > 0 $ which formally gives $ n \in ( - \infty, 0) \cup (\frac{1}{2}, 1) \cup (\frac{5}{4}, \infty) $, $ \frac{\omega_{0} (1+2n-2n^{2})}{(2n-1)(4n-5)} > 0 $ and $ \tilde{\varphi_{0}} < 0 $, as well as $ \tilde{w_{0}} < 0 $ and $ \tilde{w_{0}} \tilde{\varphi_{0}} <  2 $. 

For $ n \in ( - \infty, \frac{1-\sqrt{3}}{2}) \cup (\frac{1}{2}, 1) \cup (\frac{1+\sqrt{3}}{2}, \infty)  $ we have two cases:
\ba
C_{0} & > & 0, \,\, 2 e^{- \frac{\varphi_{0}}{2 C_{0} }} < r <  2 e^{- \frac{\varphi_{0} - M}{2 C_{0} }} \,\,\,\,
\rm{and}
\nonumber \\ 
C_{0} & < & 0, \,\, 2 e^{ \frac{\varphi_{0} - M}{2 |C_{0}| }} < r <  2 e^{ \frac{\varphi_{0}}{2 |C_{0}| }}, \,\,\,  M = \Big( \frac{(2n-1)(4n-5)}{\omega_{0} (1+2n-2n^{2})}\, \big(\frac{2}{ |\tilde{w_{0}} |} - |\tilde{\varphi_{0}}| \big) \Big)^{ - \frac{1+2n-2n^{2}}{(2n-1)(4n-5)}}.
\ea
As $ r \longrightarrow 2 e^{\mp \frac{\varphi_{0}}{2 |C_{0}| }}$ ($ C_{0} \gtrless 0 $) the functions $ F $ and $ \psi {\omega}^2 - \alpha^{2} \psi^{-1} $ tend to $ \infty $ and $ \psi \longrightarrow - \infty $. For $ n \in ( \frac{1-\sqrt{3}}{2}, 0) \cup (\frac{5}{4}, \frac{1+\sqrt{3}}{2})  $ we also have two cases:
\ba
C_{0} & > & 0, \,\,  r >  2 e^{- \frac{\varphi_{0} - \frac{1}{M}}{2 C_{0} }} \,\,\,\,
\rm{and}
\nonumber \\ 
C_{0} & < & 0, \,\, 0 < r <  2 e^{ \frac{\varphi_{0} - \frac{1}{M}}{2 |C_{0}| }}, 
\ea
wherein $ F \longrightarrow 0 $, $ \psi \longrightarrow - \infty  $ and $ \psi {\omega}^2 - \alpha^{2} \psi^{-1} \longrightarrow \infty $ as $ x \longrightarrow \infty $ ($ C_{0} >  0 $) and $ F \longrightarrow \infty $, $ \psi \longrightarrow - \infty  $ and $ \psi {\omega}^2 - \alpha^{2} \psi^{-1} \longrightarrow \infty $ as $ x \longrightarrow 0 $ ($ C_{0} <  0 $).


\section{Conclusions}

In this paper we have found exact vacuum solutions for metric power-law $f(R)$-gravity model in four-dimensional space-time in which all functions depend on combinations of two coordinates. These solutions seem to be new. Now we consider non-diagonal metrics (diagonal ones are considered in particular in the cited papers, as well as in \cite{MSh}). Not all of the solutions obtained are interesting from the physical point of view, some solutions are formal; the metrics they describe cannot be reasonably interpreted. In this article we present solutions that satisfy the correct signature and may be of most interest. 

To exactly integrate $ R^{n} $-theory equations we consider ansatz (11) that excludes the cases $ n = 1 $ and $ n = 2 $. Hence it follows that the case of General Theory of Relativity is not contained in the considered model. Therefore, it is not surprising that the scalar curvature may tend to zero while the metric does not become flat and the components of the metric tensor may even tend to infinity or zero. 

Despite the fact that in the introduced variables (17)-(19) the equations of the theory are integrated in quadratures, it is possible to obtain solutions as explicit functions of these variables only for some $ n $. It is noteworthy that most of the values of $ n $ just coincide with the boundaries of the intervals of stable and unstable nodes in the phase space studied in the works \cite{CDCT2005}-\cite{CB2006_2} and \cite{LCD2006}.

Since the $f(R) \sim R^n$-theory continues to attract the attention of researchers, it seems interesting to find and study exact expressions for the metrics that arise within this theory. We carried out an elementary analysis and presented the graphs of the obtained solutions which in our opinion are the easiest to interpret, for example, solutions that depend only on the radial variable. This can help highlight the most physically significant solutions to further explore the properties of the space-time configurations they describe. Therefore this article suggests further development. One can try to abandon the chosen ansatz and solve the equations of the theory for arbitrary $ n $. Also solutions of equations (33) were not studied in this paper; this is planned to be done in the near future.

\end{document}